\PassOptionsToPackage{unicode}{hyperref}
\PassOptionsToPackage{hyphens}{url}
\PassOptionsToPackage{dvipsnames,svgnames,x11names}{xcolor}
\documentclass[
  12pt,
]{article}
\usepackage{xcolor}
\usepackage[left=1in,right=1in,top=1in,bottom=1.25in]{geometry}
\usepackage{amsmath,amssymb}
\usepackage{iftex}
\ifPDFTeX
  \usepackage[T1]{fontenc}
  \usepackage[utf8]{inputenc}
  \usepackage{textcomp} 
\else 
  \usepackage{unicode-math} 
  \defaultfontfeatures{Scale=MatchLowercase}
  \defaultfontfeatures[\rmfamily]{Ligatures=TeX,Scale=1}
\fi
\usepackage{lmodern}
\ifPDFTeX\else
\fi
\IfFileExists{upquote.sty}{\usepackage{upquote}}{}
\IfFileExists{microtype.sty}{
  \usepackage[]{microtype}
  \UseMicrotypeSet[protrusion]{basicmath} 
}{}
\usepackage{setspace}
\usepackage{graphicx}
\makeatletter
\newsavebox\pandoc@box
\newcommand*\pandocbounded[1]{
  \sbox\pandoc@box{#1}%
  \Gscale@div\@tempa{\textheight}{\dimexpr\ht\pandoc@box+\dp\pandoc@box\relax}%
  \Gscale@div\@tempb{\linewidth}{\wd\pandoc@box}%
  \ifdim\@tempb\p@<\@tempa\p@\let\@tempa\@tempb\fi
  \ifdim\@tempa\p@<\p@\scalebox{\@tempa}{\usebox\pandoc@box}%
  \else\usebox{\pandoc@box}%
  \fi%
}
\def\fps@figure{htbp}
\makeatother
\usepackage[]{natbib}
\usepackage{inconsolata} \usepackage{float} \usepackage{caption} \usepackage{booktabs} \usepackage{makecell} \usepackage{multirow}   \usepackage{relsize}
\usepackage{bookmark}
\IfFileExists{xurl.sty}{\usepackage{xurl}}{} 
\hypersetup{
  colorlinks=true,
  linkcolor={magenta},
  filecolor={Maroon},
  citecolor={magenta},
  urlcolor={magenta},
  pdfcreator={LaTeX via pandoc}}

\title{\textscale{0.9}{A Bayesian circular mixed-effects model for explaining variability in directional movement in American football}}
\author{\textscale{0.9}{Quang Nguyen}
\qquad \qquad \qquad \textscale{0.9}{Ronald Yurko}\\
\strut \\
\textscale{0.9}{Department of Statistics \& Data Science}\\
\textscale{0.9}{Carnegie Mellon University}\\
\textscale{0.9}{Pittsburgh, PA 15213}\\
\textscale{0.9}{\texttt{\{quang,\,ryurko\}@stat.cmu.edu}}}
\date{\hfill\break
\textscale{0.9}{July 8, 2025}\\}

\begin{document}
\maketitle
\begin{abstract}
Change of direction is a key element of player movement in American
football, yet there remains a lack of objective approaches for in-game
performance evaluation of this athletic trait. Using tracking data, we
propose a Bayesian mixed-effects model with heterogeneous variances for
assessing a player's ability to make variable directional adjustments
while moving on the field. We model the turn angle (i.e., angle between
successive displacement vectors) for NFL ball carriers on both passing
and rushing plays, focusing on receivers after the catch and running
backs after the handoff. In particular, we consider a von Mises
distribution for the frame-level turn angle and explicitly model both
the mean and concentration parameters with relevant spatiotemporal and
contextual covariates. Of primary interest, we include player random
effects that allow the turn angle concentration to vary by ball carrier
nested within position groups. This offers practical insight into player
evaluation, as it reveals the shiftiest ball carriers with great
variability in turning behavior. We illustrate our approach with results
from the first nine weeks of the 2022 NFL regular season and explore
player-specific and positional differences in turn angle variability.\\
\strut 
\end{abstract}

\setstretch{1.2}
\newpage
\captionsetup{font={stretch=1.2}}
\renewcommand{\arraystretch}{1.2}

\section{Introduction}\label{introduction}

\subsection{Background: change of direction in American
football}\label{background-change-of-direction-in-american-football}

American football is a dynamic and intricate sport due to the complexity
of many athletes interacting over the course of every play. Player
movement in football consists of a variety of elements that together
contribute to athletic performance on the field. One vital component of
player movement is change of direction, which demonstrates the ability
of a player to rapidly and efficiently alter their trajectory. On
offense, players use this movement skill to create separation and evade
tackles---whether it is a running back cutting through traffic, a
receiver breaking off a route, or a quarterback escaping pressure. On
defense, change of direction enables defenders to close space quickly
and maintain leverage against the opposing players on offense.

In the National Football League (NFL), the most commonly-known way to
assess change of direction is through timed physical drills at the NFL
Scouting Combine, namely the three-cone and short shuttle. In a
three-cone drill (or L-drill), a player is asked to run, stop, and loop
around three cones arranged in a right angle ``L'' shape. The short
shuttle, also known as the 20-yard shuttle or 5--10--5 drill, is a test
that requires an individual to first sprint five yards to one side,
reverse and sprint ten yards the opposite way, then run five yards back
to the starting point. Both of these tests are designed to evaluate
agility, quickness, and lateral movement of NFL prospects. These
traditional drills, while providing some baseline for evaluating change
of direction, have apparent limitations. Specifically, they fail to
capture how players respond in live, real-time scenarios where movement
is influenced by spatial constraints and different moving pieces on the
field.

As a side note, there are other timed drills in sports science related
to the aforementioned NFL Combine evaluations. These include the T-test,
505, Illinois, and Arrowhead agility tests, to name a few. As an
example, the 505 test focuses on 180-degree turning performance and
involves a 15-meter sprint, followed by a 180-degree turn and a 5-meter
sprint back. Based on this test, one can accordingly calculate the
so-called change of direction deficit by computing the difference
between the 505 time and the time it takes for a straight-line sprint of
an equivalent distance. We also note a substantial number of articles in
the sports science literature that analyze change of direction from a
performance training perspective (see \citet{brughelli2008understanding}
for a thorough review). Within this body of work, the goal is usually to
link change of direction with factors like strength, conditioning,
fatigue, or injury risk. Recently, \citet{giles2019machine} and
\citet{giles2024quantifying} use camera tracking data to detect and
classify directional movement in tennis. Since all of the above
contributions do not attempt to model in-game change of direction
performance of athletes, they are of limited relevance to our paper.

Despite its importance, there are currently no widely-adopted
\emph{public} approaches for evaluating directional movement in American
football using in-game data, to the best of our knowledge\footnote{We
  are aware that NFL teams and vendors do have proprietary methods for
  evaluating in-game change of direction. For example, Teamworks
  Intelligence uses ideas from the aforementioned change of direction
  deficit concept in sports science to develop their own metric
  \citep{kassam2025minding}.}. However, with the increasing availability
of player tracking data, there are ample opportunities to quantify
movement traits like change of direction directly from observed on-field
information. Below, we provide a brief summary of recent developments in
football analytics with tracking data.

\subsection{Background: player tracking data in American
football}\label{background-player-tracking-data-in-american-football}

The advent of player tracking data has transformed the field of sports
analytics in recent years \citep{baumer2023big, kovalchik2023player}. In
the NFL, tracking data are collected via the Next Gen Stats system with
radio frequency identification (RFID) chips placed inside the football
and shoulder pads of players. This outputs spatial location and
trajectory information for all 22 players on the field during any given
play at a rate of 10 Hz (i.e., 10 frames per second). The granularity
and continuous nature of tracking data enable in-depth exploration of
various topics in American football that were not previously possible
with traditional box-score statistics or play-level data.

In an attempt to crowdsource public insight into player tracking data,
the NFL organizes an annual contest called the Big Data Bowl
\citep{lopez2020bigger}. Every year since 2019, the NFL releases a
sample of player tracking data to accompany a football-related Big Data
Bowl theme such as pass coverage, special teams, linemen, tackling, and
pre-snap motion. The competition has consequently established the
academic literature on NFL tracking data, promoting data-driven
innovations in football analytics \citep{lopez2020bigger}. Recent
developments in this area include models for hypothetical completion
probability \citep{deshpande2020expected}, receiver route detection
\citep{chu2020route}, pass coverage identification
\citep{dutta2020unsupervised}, and within-play expected points
\citep{yurko2020going}; as well as novel performance metrics for pass
rush pressure \citep{nguyen2024here}, tackling
\citep{nguyen2025fractional}, and snap timing
\citep{nguyen2025multilevel}. It is important to note that some of the
aforementioned contributions offer insights into previously understudied
areas of American football like defensive performance and pre-snap
actions.

\subsection{Our contribution}\label{our-contribution}

In this paper, we introduce a modeling framework to assess in-game
change of direction ability of players in American football. We first
characterize change of direction at the frame level using the concept of
turn angle, borrowing from the animal movement literature. This simply
measures the angular deviation between consecutive movement steps of NFL
players during any given play. Our primary quantity of interest is the
variability in turn angle, which illustrates the level of consistency in
directional adjustment a player displays across different moments in
time. We hypothesize that higher variability in turn angle is beneficial
from a performance evaluation point of view, as this may be indicative
of the ability to perform shifty movement with more sudden,
unpredictable turns.

To achieve this, we leverage fine-grained tracking data and develop a
circular mixed-effects model for player directional movement in the NFL.
Here, we focus on ball carriers on both run and pass
plays---specifically, running backs after the handoff and receivers
after the catch. We assume a von Mises distribution for the frame-level
turn angle and simultaneously model both its mean and concentration
parameter. Our framework controls for a variety of tracking and
contextual features; and most importantly, ball carrier random effects
at the concentration level, allowing us to estimate the differences in
turn angle variability among NFL players. We implement our approach in a
Bayesian setting, which naturally provides uncertainty quantification
for all model parameters and enables position-level pooling. We
highlight that our modeling technique can be extended to the evaluation
of other positions in football, including kick and punt returners on
special teams, quarterbacks on rushing plays, as well as defensive
players.

\section{Methods}\label{methods}

\subsection{Player tracking data}\label{player-tracking-data}

Our analysis relies on data from the NFL Big Data Bowl 2025
\citep{lopez2024nfl}, which comprise of spatial tracking data from the
first nine weeks of the 2022 NFL regular season, along with game, play,
and player-level supplementary information. Of primary interest, the
tracking data provide attributes on two-dimensional coordinates, speed,
direction, and orientation of every player on the field at a rate of 10
frames per second. Table \ref{tab:tracking} illustrates the typical
structure of football tracking data for an individual player during a
play. Note that besides the location and trajectory variables, the data
also contain event annotations (e.g., ball snap, handoff, tackle, etc.)
for relevant frames within each play.

\begin{table}[t]
\caption{Example of tracking data for a play during the Carolina Panthers versus Cincinnati Bengals NFL game on November 6, 2022. The data presented here are for Bengals running back Joe Mixon, and the frames included are between the ball snap and when the tackle is made. The data attributes include frame identifier (frameId), two-dimensional coordinates (x and y), speed (s), acceleration (a), distance traveled from previous frame (dis), orientation (o), direction (dir), and event tag for each frame (event). \label{tab:tracking}}
\centering
\begin{tabular}{rrrrrrrrl}
\hline \addlinespace[0.25ex]
frameId & x & y & s & a & dis & o & dir & event \\ \addlinespace[0.25ex]
\hline \addlinespace[0.25ex]
6 & 95.45 & 23.63 & 0.25 & 2.05 & 0.02 & 76.69 & 127.17 & ball\_snap \\ \addlinespace[0.25ex]
7 & 95.49 & 23.61 & 0.66 & 3.50 & 0.05 & 77.79 & 117.14 & -- \\ \addlinespace[0.25ex]
\vdots & \vdots & \vdots & \vdots & \vdots & \vdots & \vdots & \vdots & \vdots \\ \addlinespace[0.25ex]
16 & 98.32 & 23.00 & 5.15 & 2.13 & 0.50 & 89.12 & 100.99 & handoff \\ \addlinespace[0.25ex]
\vdots & \vdots & \vdots & \vdots & \vdots & \vdots & \vdots & \vdots & \vdots \\ \addlinespace[0.25ex]
25 & 102.81 & 21.10 & 5.33 & 1.75 & 0.54 & 77.30 & 123.03 & first\_contact \\ \addlinespace[0.25ex]
\vdots & \vdots & \vdots & \vdots & \vdots & \vdots & \vdots & \vdots & \vdots \\ \addlinespace[0.25ex]
43 & 107.49 & 16.59 & 2.92 & 1.73 & 0.29 & 269.27 & 137.37 & -- \\ \addlinespace[0.25ex]
44 & 107.65 & 16.42 & 2.45 & 2.81 & 0.24 & 266.75 & 136.04 & tackle \\ \addlinespace[0.25ex]
\hline
\end{tabular}
\end{table}

In this work, we study change of direction for NFL ball carriers on both
passing and running plays. For each play, we extract only frames within
a so-called \emph{ball carrier sequence} using the provided event labels
in the tracking data. In particular, for running plays, we solely focus
on running backs and consider the time window between the handoff and
the end-of-play event---namely, tackle, out-of-bound, touchdown, or
fumble. As for passing plays, we focus on wide receivers, tight ends,
and running backs after the catch. Specifically, we identify the
beginning of the ball carrier sequence as when the player catches the
football, and the end as one of the outcomes similar to what we consider
for run plays.

After preprocessing, our final dataset contains 9,480 total plays across
all 136 games played during weeks 1 through 9 of the 2022 NFL season.
Among these plays, we end up with 5,431 rushing attempts, all by running
backs. There are also 4,049 passing plays, including 2,074 with wide
receivers, 1,061 with running backs, and 914 with tight ends as the ball
carrier after the catch.

\subsection{Feature engineering}\label{sec:features}

Using the provided tracking data, we construct meaningful features for
fitting our directional movement model. Here, we aim for a simple set of
covariates that summarizes the ball carrier's location and trajectory,
as well as their relationship with other players on both offense and
defense throughout a play. We use the same strategy as
\citet{yurko2020going} and \citet{yurko2024nfl} to derive the dynamic
within-play features of interest.

\begin{table}[tbp]
\caption{List of features for the turn angle model described in Section \ref{sec:model}. All features are frame-level tracking data features, unless otherwise noted.}
\label{tab:features}
\centering
\begin{tabular}{p{0.59\textwidth}p{0.08\textwidth}p{0.1\textwidth}p{0.02\textwidth}p{0.02\textwidth}}
\hline \addlinespace[0.25ex]
\multirow{2}{*}{Feature} & \multicolumn{2}{c}{$\text{Player }$} & \multicolumn{2}{c}{Model} \\ \addlinespace[0.25ex]
& Ball carrier & Closest defender & $\mu$ & $\kappa$ \\ \addlinespace[0.25ex]
\hline \addlinespace[0.25ex]
Previous turn angle (radians) & \checkmark & & \checkmark & \\ \addlinespace[0.25ex]
Horizontal yards from target endzone & \checkmark & \checkmark & \checkmark & \\ \addlinespace[0.25ex]
Vertical yards from center of the field with respect to target endzone (positive: left side; negative: right side) & \checkmark & \checkmark & \checkmark & \\ \addlinespace[0.25ex]
Horizontal yards from first down line & \checkmark & & \checkmark & \\ \addlinespace[0.25ex]
Number of defenders in front (high: more in front; low: more behind) & \checkmark &  & \checkmark & \\ \addlinespace[0.25ex]
Number of defenders to the left (high: more on left side; low: more on right side) & \checkmark &  & \checkmark & \\ \addlinespace[0.25ex]
Number of teammates in front (high: more in front; low: more behind) & \checkmark &  & \checkmark & \\ \addlinespace[0.25ex]
Number of teammates to the left (high: more on left side; low: more on right side) & \checkmark &  & \checkmark & \\ \addlinespace[0.25ex]
Defender's speed (yards/second) & & \checkmark  & \checkmark & \\ \addlinespace[0.25ex]
Angle of motion relative to ball carrier (radians) & & \checkmark  & \checkmark & \\ \addlinespace[0.25ex]
Horizontal yards relative to ball carrier & & \checkmark  & \checkmark & \\ \addlinespace[0.25ex]
Vertical yards (absolute value) relative to ball carrier  & & \checkmark  & \checkmark & \\ \addlinespace[0.25ex]
Distance from ball carrier (yards) & & \checkmark  & \checkmark & \\ \addlinespace[0.25ex]
Speed (yards/second) & \checkmark &  & & \checkmark \\ \addlinespace[0.25ex]
Acceleration (yards/second$^2$)& \checkmark &  & & \checkmark \\ \addlinespace[0.25ex]
Cumulative distance covered (yards) & \checkmark & & & \checkmark  \\ \addlinespace[0.25ex]
Play type$^*$ (run or pass) & \checkmark  &  &  & \checkmark \\ \addlinespace[0.25ex]
Position$^*$ (RB, TE, or WR) & \checkmark & &  & \checkmark \\ \addlinespace[0.25ex]
Player varying intercept$^\dagger$ (grouped by position) &  \checkmark & &  & \checkmark \\ \addlinespace[0.25ex]
\hline
\multicolumn{5}{l}{\footnotesize $^*$Play context}\\ \addlinespace[-0.5ex]
\multicolumn{5}{l}{\footnotesize $^\dagger$Random effect}\\
\end{tabular}
\end{table}

\begin{figure}[ptb]

{\centering \includegraphics{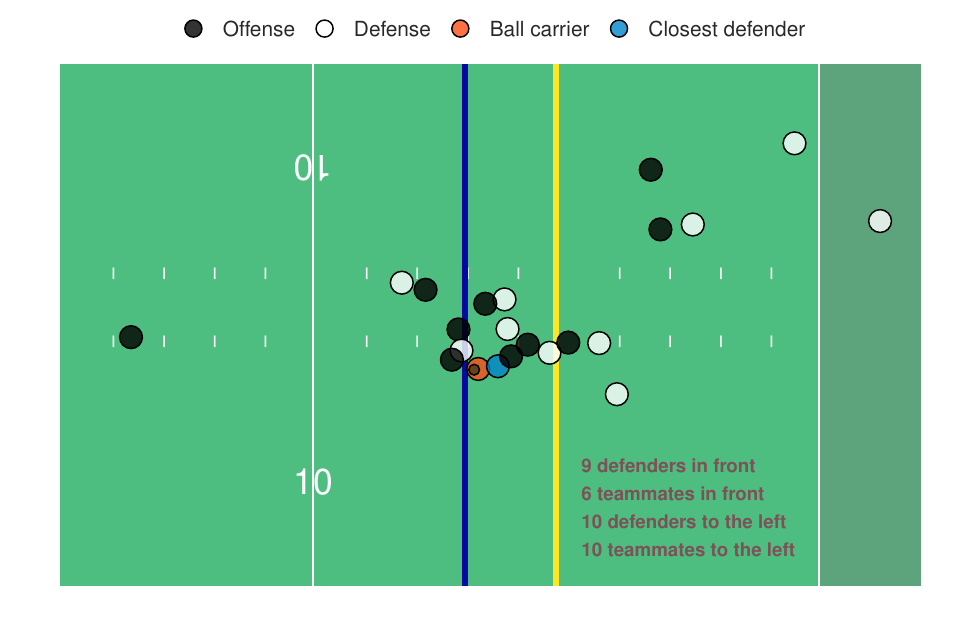} 

}

\caption{Snapshot of a given frame (obtained from tracking data) within a play during the Carolina Panthers (defense, in black) versus Cincinnati Bengals (offense, in white) NFL game on November 6, 2022. We construct dynamic tracking data features at this frame for three groups: ball carrier (highlighted in orange), closest defender (highlighted in blue), and simple counts for other players on the field with respect to the ball carrier (annotated in maroon).}\label{fig:dynamic_features}
\end{figure}

Table \ref{tab:features} provide a full list of our tracking data
features, which can be categorized into three groups (see also Figure
\ref{fig:dynamic_features} for a visual explanation). First, we consider
different positional and trajectory information about the ball carrier,
including their speed, acceleration, distance covered, horizontal
displacement relative to the target end zone and first down line, and
lateral displacement with respect to the center of the field. We also
create frame-level features describing the closest defender within a
play. These include the distance and angle between the nearest defender
and ball carrier, as well as the defender's speed and location on the
field. Moreover, we obtain simple counts such as the number of defenders
and teammates on each side and in front of the ball carrier at any given
frame. We emphasize that this set of covariates is only a starting
point, and it is certainly possible to consider more extensive and
refined tracking data features. We choose this simple representation of
the spatial relationships between players on the field for ease of model
fitting, as these features are treated as linear fixed effects in our
modeling framework described in Section \ref{sec:model}.

\subsection{Measuring change of
direction}\label{measuring-change-of-direction}

\begin{figure}[ptb]

{\centering \includegraphics{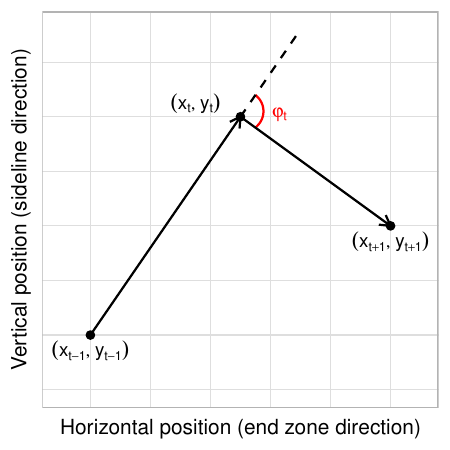} 

}

\caption{Geometric representation of the instantaneous turn angle $\varphi_t$, computed at frame $t$ using successive player positions $(x_{t-1}, y_{t-1})$, $(x_t, y_t)$, and $(x_{t+1}, y_{t+1})$. Axes reflect the standardized football field coordinates, with the horizontal and vertical axes aligned to the end zone and sideline directions, respectively.}\label{fig:turn_angle_diagram}
\end{figure}

To measure change of direction in American football, we draw inspiration
from the animal movement literature (see \citet{hooten2017animal} for a
complete overview) and use the following characterization. For a player
at each time point within a play, we compute the \textbf{turn angle},
which refers to the angle a player takes between consecutive steps,
i.e., the angle between consecutive displacement vectors. Formally, let
\((x_t, y_t)\) be the observed player coordinates at frame \(t\) within
a play, for \(t=1, \dots, T + 1\). For an individual player at frame
\(t\), the turn angle \(\varphi_t\) is the change in bearing between two
successive time intervals \([t-1, t]\) and \([t, t + 1]\). That is,
\[\varphi_t = b_t - b_{t-1},\] where the bearing angle
\(b_t \in [-\pi, \pi]\) is defined as
\[b_t = \text{atan}2 (y_{t+1} - y_t, x_{t+1} - x_t),\] with
\(\text{atan}2(\cdot)\) representing the two-argument inverse tangent
function. In Figure \ref{fig:turn_angle_diagram}, we display a visual
explanation of our turn angle metric.

\begin{figure}[ptb]

{\centering \includegraphics{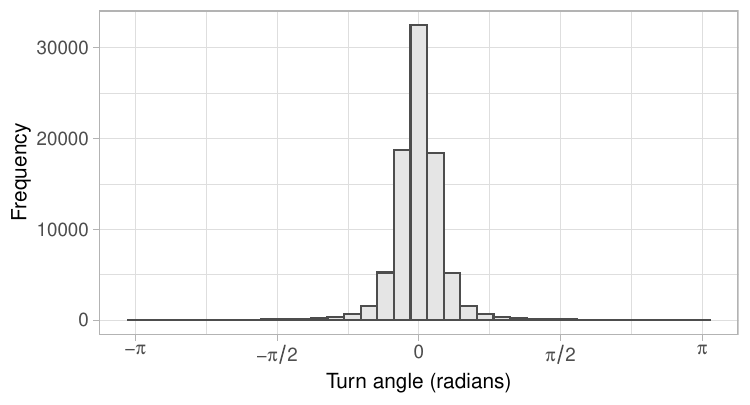} 

}

\caption{Distribution of turn angle for NFL ball carriers during the first nine weeks of the 2022 regular season.}\label{fig:turn_angle_distribution}
\end{figure}

For context, Figure \ref{fig:turn_angle_distribution} displays the turn
angle distribution for the considered ball carriers during the first
nine weeks of the 2022 NFL season. We notice that the turn angle values
are highly concentrated around zero. This suggests that most players
exhibit relatively small directional changes between successive frames
across the considered plays.

\subsection{Turn angle model}\label{sec:model}

Let \(\varphi_{ijt}\) denote the turn angle for ball carrier \(j\) at
frame \(t\) of play \(i\). To understand the typical turning direction
for NFL players and how variable their directional movement are, we fit
the following circular mixed-effects model: \[
\begin{aligned}
\varphi_{ijt} &\sim \textsf{von Mises}(\mu_{ijt}, \kappa_{ijt}),\\
\tan \frac{\mu_{ijt}}{2} &= \alpha_0 + \mathbf{x}_{ijt}^\top \boldsymbol{\beta},\\
\log \kappa_{ijt} &= \gamma_0 + \mathbf{z}_{ijt}^\top \boldsymbol{\psi} + u_j,\\
u_j &\sim \mathcal N\left(0, \sigma^2_{p[j]}\right).\\
\end{aligned}
\]

In this model, we use a von Mises distribution for the response
\(\varphi_{ijt}\), parameterized in terms of its mean \(\mu\) and the
concentration parameter \(\kappa > 0\). This is a popular distributional
choice for modeling an angular outcome variable. Here, \(\kappa\)
denotes the concentration parameter of a von Mises distribution,
illustrating a reciprocal measure of variability. In other words, a
higher \(\kappa\) value corresponds to a more concentrated (i.e., less
dispersed) turn angle distribution. Note that in this specification, we
explicitly model both the mean \(\mu\) and concentration \(\kappa\)
parameters. This offers information about not only which direction
players tend to move, but also the variability in how they do so.

First, the mean parameter \(\mu\) is modeled with a tan-half link
function, restricting the turn angle values to \((-\pi, \pi)\). We
account for tracking data features describing the ball carriers and
other players on the field (as summarized in Section \ref{sec:features}
and Table \ref{tab:features}) through \(\mathbf{x}_{ijt}\), and estimate
the coefficients \(\boldsymbol{\beta}\) as fixed effects. It is worth
noting that the turn angle at the previous frame \(\varphi_{ij,t-1}\) is
included as part of our covariate information when modeling the mean
turn angle at frame \(t\). This encodes the notion of directional
persistence, capturing player tendency to make successive turns in a
similar manner.

Next, in the model for the concentration parameter \(\kappa\) (using a
log link function), we condition on various movement and contextual
features \(\mathbf{z}_{ijt}\) (with corresponding coefficients
\(\boldsymbol{\psi}\)) that likely contribute to the variability in turn
angle (see Table \ref{tab:features}). Specifically, we adjust for the
ball carrier's speed, as a player moving very fast may be limited in how
sharply they can turn, whereas larger turns can be executed more easily
at a slower speed. In addition, we control for the frame-level
acceleration, since high acceleration or deceleration may signal a
change in a player movement state (e.g., making cuts). We also include
the cumulative distance covered, showing how much a player has moved
during the play, which might reflect fatigue or time spent maneuvering
in open space. As for contextual information, we take into account the
play type (pass or run) and position of the ball carrier (running backs,
tight ends, or wide receivers). Both of these covariates are encoded as
indicator variables, with pass plays and running backs being the
reference levels for the play type and position, respectively.

Of primary interest, we include a random intercept term \(u_j\) for ball
carrier \(j\), with variance \(\sigma^2_{p[j]}\) depending on player
position. Here, \(p_{[j]}\) denotes the position group of ball carrier
\(j\), so each position (running backs, wide receivers, and tight ends)
has its own variance. This measures how much ball carriers within a
given position differ in their directional movement variability. By
including the ball carrier random effect \(u_j\), we aim to infer how
consistent or variable a player's turning behavior is, relative to the
average ball carrier in the same position group. Ultimately, this
reveals a player's ability to exhibit more variable directional
adjustment across different time points, as it is reasonable to expect
some ball carriers to be shiftier than others. To some extent, having
higher variability in turn angle may demonstrate an athletic trait that
can be useful from a performance evaluation standpoint.

We implement our turn angle model using a Bayesian approach with
\texttt{Stan} \citep{carpenter2017stan} via the \texttt{brms} package in
\texttt{R} \citep{burkner2017brms, r2025language}. This provides natural
uncertainty quantification for all model parameters, whose posterior
distributions are estimated using MCMC via the no-U-turn sampler
\citep{hoffman2014nuts}. We choose weakly informative prior
distributions for the parameters in our turn angle model. In particular,
we specify vague \(\text{half-}t_3\) priors (i.e., a Student's \(t\)
distribution with 3 degrees of freedom and restricted to be
non-negative) for the standard deviation parameters \(\sigma_p\)
\citep{gelman2006prior}.

To fit this model, we use 4 parallel chains, each with 3,500 iterations
and a burn-in of 1,500 draws. This yields 8,000 samples for posterior
analysis of the model estimates. To assess convergence of the MCMC
procedure, we perform visual checks of trace plots and compute the
\(\hat R\) statistic \citep{gelman1992inference}. We find good evidence
of convergence as revealed by the trace plots and all \(\hat R\) values
being close to 1. We also observe no issues with model efficiency based
on the effective sample size \citep{gelman2013bayesian} for each
parameter.

\section{Results}\label{results}

In this section, we investigate the results of fitting our Bayesian
circular mixed-effects model for the frame-level turn angle described in
Section \ref{sec:model}. Although the mean turn angle (\(\mu\)) model
captures the tendency of directional movement, we are more interested in
gaining insights into the variability in player turning behavior in the
NFL. Thus, in the analysis that follows, we focus on our estimates when
modeling the turn angle concentration \(\kappa\).

\subsection{Fixed effects of turn angle concentration
model}\label{fixed-effects-of-turn-angle-concentration-model}

Table \ref{tab:fixedeff} shows the estimated coefficients of the fixed
effect terms in our turn angle concentration model. We first explore the
relationships between the variability in change of direction and the
explanatory movement variables---speed, acceleration, and cumulative
distance traveled.

Here, we see that the concentration in directional movement increases as
speed increases (\(\hat \psi_{\textsf{s}} = 0.709\), 95\% credible
interval: \([0.705, 0.713]\)). In other word, faster ball carriers tend
to move in more consistent directions. This indicates that it is more
difficult for players to make sharp cuts when carrying the football at
their top speed. Our model also reveals a negative relationship between
acceleration and turn angle concentration
(\(\hat \psi_{\textsf{a}} = -0.094\), 95\% credible interval:
\([-0.101,-0.087]\)). Specifically, higher acceleration is associated
with less turning consistency, or more variable turning behavior. This
matches with what we typically see on the football field, as high
acceleration events (e.g., players cutting to avoid a tackle or breaking
into open space) often involve sudden directional adjustments in the
ball carrier movement path. These are indeed the cases where we would
expect more variability in change of direction. Moreover, we observe
that as a ball carrier covers more distance, their turning behavior
slightly becomes less predictable
(\(\hat \psi_{\textsf{dis}} = -0.015\), 95\% credible interval:
\([-0.016,-0.014]\)). This perhaps suggests that players may adjust
their pace and direction more erratically in later moments of a play.

As for play context, it appears that players on run plays (during the
ball carrier sequence) demonstrate higher turn angle concentration than
those on pass plays (\(\hat \psi_{\text{run}} = 0.044\), 95\% credible
interval: \([0.011, 0.078]\)). This makes intuitive sense, as ball
carriers after the handoff on run plays tend to follow more structured
trajectories, compared to receivers after the catch who often have to
improvise in open field. Regarding positional effects, we find that wide
receivers and running backs differ in turn angle concentration. In
particular, wide receivers display less consistency (i.e., more
variation) in change of direction, as indicated by
\(\hat \psi_{\textsf{WR}} = -0.134\) (95\% credible interval:
\([-0.200, -0.065]\)). This ties into our above analysis of how play
type relates to turn angle concentration, since for pass plays we only
focus on frames after the catch. Additionally, although the coefficient
estimate for tight ends is positive
(\(\hat \psi_{\textsf{TE}} = 0.064\)), we are not certain whether tight
ends exhibit more concentrated directional movement than running backs,
since the corresponding 95\% credible interval of \([-0.021, 0.151]\)
includes zero.

\begin{table}[tbp]
\caption{Posterior estimates for the fixed effect coefficients $\boldsymbol \psi$ when modeling the turn angle concentration $\kappa$. Note that the reference play level is pass play, denoted \textsf{play:pass}, and the reference position level is running backs, denoted \textsf{position:RB}.\label{tab:fixedeff}}
\centering
\begin{tabular}{lrrrr}
\hline \addlinespace[0.25ex]
\multicolumn{1}{c}{} & \multicolumn{2}{c}{Posterior summary} & \multicolumn{2}{c}{95\% credible interval} \\ \addlinespace[0.25ex]
& Mean & SD & Lower & Upper \\ \addlinespace[0.25ex]
\hline \addlinespace[0.25ex]
Intercept & $1.309$ & $0.026$ & $1.258$ & $1.361$ \\ \addlinespace[0.25ex]
Speed & $0.709$ & $0.002$ & $0.705$ & $0.713$ \\ \addlinespace[0.25ex]
Acceleration & $-0.094$ & $0.004$ & $-0.101$ & $-0.087$ \\ \addlinespace[0.25ex]
Distance covered & $-0.015$ & $0.001$ & $-0.016$ & $-0.014$ \\ \addlinespace[0.25ex]
$I_{\textsf{\{play:run\}}}$ & $0.044$ & $0.017$ & $0.011$ & $0.078$ \\ \addlinespace[0.25ex]
$I_{\textsf{\{position:TE\}}}$ & $0.064$ & $0.043$ & $-0.021$ & $0.151$ \\ \addlinespace[0.25ex]
$I_{\textsf{\{position:WR\}}}$ & $-0.134$ & $0.034$ & $-0.200$ & $-0.065$ \\ \addlinespace[0.25ex]
\hline
\end{tabular}
\end{table}

\subsection{Positional differences in random effect
variance}\label{positional-differences-in-random-effect-variance}

Next, we examine the estimates for the position-specific standard
deviation parameters \(\sigma_p\) in our turn angle model (see Table
\ref{tab:randeff}). This tells us how much ball carriers within a given
position vary in their turn angle concentration. Our results suggest
that wide receivers (\(\hat \sigma_{\textsf{WR}} = 0.304\), 95\%
credible interval: \([0.257, 0.355]\)) and tight ends
(\(\hat \sigma_{\textsf{TE}} = 0.300\), 95\% credible interval:
\([0.234, 0.378]\)) have higher position-specific standard deviation
estimates than running backs (\(\hat \sigma_{\textsf{RB}} = 0.135\),
95\% credible interval: \([0.111, 0.164]\)). This appears reasonable,
given the different roles and types of movement that players in those
position groups typically perform.

\begin{table}[tbp]
\caption{Posterior estimates for the standard deviation of the ball carrier concentration random effect (grouped by position) in our turn angle model. \label{tab:randeff}}
\centering
\begin{tabular}{lrrrr}
\hline \addlinespace[0.25ex]
\multicolumn{1}{c}{} & \multicolumn{2}{c}{Posterior summary} & \multicolumn{2}{c}{95\% credible interval} \\ \addlinespace[0.25ex]
Parameter & Mean & SD & Lower & Upper \\ \addlinespace[0.25ex]
\hline \addlinespace[0.25ex]
$\sigma_{\textsf{RB}}$ & $0.135$ & $0.014$ & $0.111$ & $0.164$ \\ \addlinespace[0.25ex]
$\sigma_{\textsf{TE}}$ & $0.300$ & $0.037$ & $0.234$ & $0.378$ \\ \addlinespace[0.25ex]
$\sigma_{\textsf{WR}}$ & $0.304$ & $0.025$ & $0.257$ & $0.355$ \\ \addlinespace[0.25ex]
\hline
\end{tabular}
\end{table}

To elaborate further, since we only consider moments after the catch for
wide receivers and tight ends, these are usually open-field,
unstructured movements with high variability. As such, some players
might have the tendency to sprint straight, while others might use
different forms of evasive maneuvers (e.g., cuts, jukes, etc.) when
carrying the football. Hence, ball carriers within both of these
positions differ considerably in their turn angle variability.

Meanwhile, running back movements are generally more structured and
system-driven, whether during post-catch moments in pass plays or
post-handoff moments in run plays. In these settings, the ball carrier
sequence typically starts behind the line of scrimmage, either with a
handoff from the quarterback to the running back, or with the running
back catching a short pass in the offensive backfield. These are often
constrained by different blocking schemes and run gaps, resulting in
more similar levels of directional movement pattern among running backs.

\subsection{Ball carrier ratings}\label{ball-carrier-ratings}

\begin{figure}[ptb]

{\centering \includegraphics{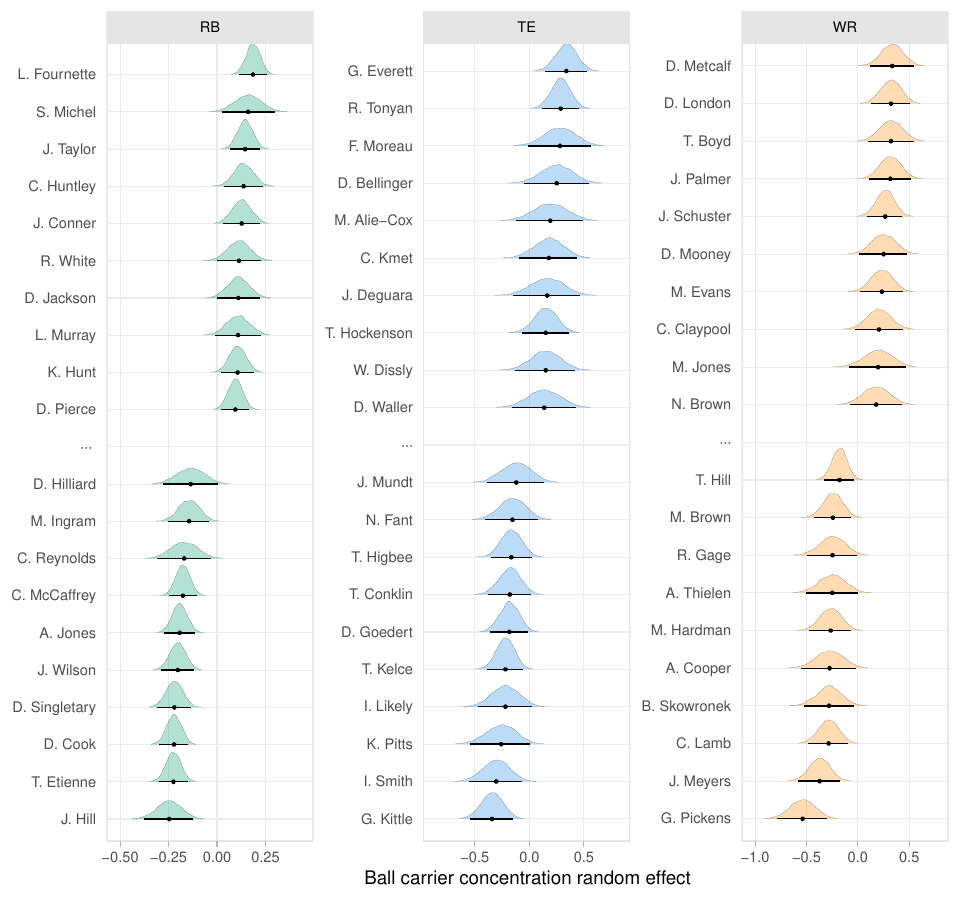} 

}

\caption{Posterior distributions of the ball carrier concentration random effect for NFL running backs (with at least 25 plays), tight ends (with at least 10 plays), and wide receivers (with at least 15 plays) over the first nine weeks of the 2022 NFL regular season. For each player, the posterior mean point estimate and corresponding 95\% credible interval are depicted. Here, \textbf{a lower posterior mean demonstrates higher variability (lower concentration) in turn angle}, and results are shown for the top and bottom 10 players within each position based on the posterior mean values.}\label{fig:player_concentration_ratings}
\end{figure}

Our Bayesian circular mixed-effects model provides straightforward
posterior summary for comparing the change of direction performance of
NFL ball carriers during the first nine weeks of the 2022 regular
season. Figure \ref{fig:player_concentration_ratings} presents posterior
distributions of the concentration random effect \(u_j\) for the three
considered position groups---running backs (with at least 25 plays),
tight ends (with at least 10 plays), and wide receivers (with at least
15 plays). Here, the players are ordered by the posterior mean
estimates, which capture their ability to exhibit variable turning
behavior on the field. More specifically, a higher posterior mean value
corresponds to lower variability (i.e., higher concentration) in making
directional changes when carrying the ball.

In our rankings, Justice Hill stands out with the most variability in
turn angle relative to the rest of the running backs. This aligns with
the eye test, as Hill is known as a shifty and agile runner, which dates
back to his college career \citep{durkin2019clock}. Our results also
show players like Leonard Fournette and Jonathan Taylor with low
variability in change of direction. Indeed, these running backs are
known to have a tendency to run straight ahead with power and speed
rather than making variable directional adjustments
\citep{kramer2019jonathan}.

Next, among the wide receivers, we notice DK Metcalf as the ball carrier
with the least variability in change of direction. This is not
surprising, as Metcalf tends to rely heavily on straight-line movement
and speed, more so than displaying erratic lateral movements
\citep{levy2019combine}. George Pickens, on the other hand, exhibits the
greatest variability in turn angle in our receiver rankings. During his
time with the Pittsburgh Steelers, Pickens was known for his ability to
perform a variety of moves to make defenders miss after the catch.

As for tight ends, our leaderboard is very much consistent with
conventional rankings of players in this position. Specifically, George
Kittle, Kyle Pitts, and Travis Kelce all belong in the top five in terms
of high variability in turn angle. This offers support to our estimates
as an effective measure of the ability to make sudden, more variable
change of direction. It is worth noting that relative to running backs
and wide receivers, we observe wider 95\% credible intervals for tight
ends. Furthermore, a number of the posterior distributions in Figure
\ref{fig:player_concentration_ratings} are not entirely above or below
zero. Evidently, there is uncertainty in our estimates, which is not
surprising due to our limited sample of considered plays.

Note that among the running backs and wide receivers, there are clear
differences in the 95\% credible intervals of the player concentration
random effects. In particular, there are no overlaps between the
intervals for the top and bottom ball carriers in these two position
groups. Thus, our estimates reliably differentiate between the players,
providing discriminative power for measuring change of direction in the
NFL.

\subsection{Speed-and-turn movement
profiles}\label{speed-and-turn-movement-profiles}

\begin{figure}[t]

{\centering \includegraphics{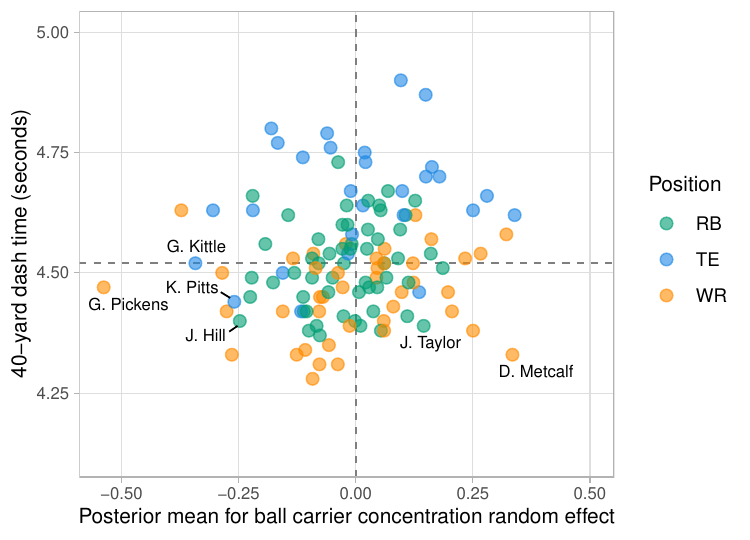} 

}

\caption{Relationship between the posterior mean of the ball carrier concentration random effect and the 40-yard dash time recorded at the NFL Scouting Combine. The horizontal dashed line represents the median 40-yard time. We observe a positive, weak overall correlation ($r=0.135$), and there are similar levels of correlation within the three position groups---running backs ($r=0.139$), tight ends ($r=0.237$), and wide receivers ($r=0.118$).}\label{fig:speed_turn_profiles}
\end{figure}

To better understand player movement profiles in American football, we
examine our turn angle variability summary alongside a speed measure for
NFL players. Figure \ref{fig:speed_turn_profiles} shows the relationship
between the posterior mean of the ball carrier concentration random
effect and their 40-yard dash time at the NFL Scouting Combine (obtained
via the \texttt{nflreadr} package \citep{ho2025nflreadr} in \texttt{R}).
The scatterplot reveals a weak positive relationship between turning
variability and speed---both overall \((r=0.135)\) and within
positions---highlighting the diversity of movement profiles among NFL
ball carriers.

We notice different player traits as revealed by the speed-and-turn
joint distribution. For instance, DK Metcalf showcases exceptional
speed, but low variability in turning pattern. This once again
emphasizes Metcalf's ability to move extremely fast with the football,
while also displaying less lateral agility. Meanwhile, George Pickens is
a shifty ball carrier who records a faster 40-yard time than the median
value. This indicates that Pickens exhibits good quickness and can also
change direction rapidly. As for running backs, players like Jonathan
Taylor are more effective at running straight with speed to gain ground
quickly, while Justice Hill showcases both great quickness and more
erratic directional movement. It also makes sense that many tight ends
appear in the top-right quadrant, given the physical makeup of this
position. With the exception of players like Kyle Pitts, most tight ends
appears to have slower speed and less variability in turning behavior.

\section{Discussion}\label{discussion}

Change of direction is an important component of player movement in
American football, but there exists limited research on in-game
evaluation of this skill. In this work, we provide the first public
approach for assessing change of direction performance of NFL players
using publicly-available football tracking data. We propose a Bayesian
mixed-effects model to evaluate an NFL ball carrier's ability to display
variable directional movement on the field. Our modeling framework is
illustrated for running backs after the handoff on run plays and
receivers after the catch on pass plays. In our model, we assume a von
Mises distribution for the frame-level turn angle, with primary interest
in its concentration parameter. We adjust for relevant tracking data and
play context features when modeling both the turn angle mean and
concentration, and random intercept for ball carrier nested within
positions at the concentration level.

Our findings suggest differences in turn angle variability across
players within position groups, with running backs having more similar
levels of concentration compared to tight ends and wide receivers. This
is an intuitive result, consistent with the typical roles and types of
movement on the field each position is involved in. Additionally, we are
able to identify top-performing players based on our variability in
change of direction measure. We also observe different ball carrier
movement profiles when examining our turn angle results together with a
player's speed measured at the NFL Combine. For instance, some players
are more effective at running straight with speed (e.g., DK Metcalf),
while others excel at making erratic, variable turns (e.g., George
Pickens).

One great benefit to our approach is that it can be extended to the
evaluation of other positions in American football. This applies to
different aspects across the board, from quarterback scrambles and
receivers \emph{before} the catch on offense, to pass rush and pass
coverage movement on defense, and even punt and kickoff returners on
special teams. We note that with access to other data sources such as
biomechanics data (or pose data), besides capturing \emph{what} a player
does with our model (e.g., variable turning), we can also understand
\emph{how} they do it (e.g., through hip or foot positioning). As an
example, defensive back movement includes elements like backpedaling,
transition into sprint, and sudden cuts or turns when reacting to
receivers. With biomechanics data, variables related to hip orientation
or foot placement can be useful for explaining directional control of
the players during pass coverage. While these data are currently not
open to the public, proprietary analysis using pose estimation has shown
promising insights into defensive back directional movement
\citep{maclean2024defensive}. We look forward to incorporating pose data
features into our framework once they become publicly available to
better understand player turning behavior in American football.

\section*{Acknowledgments}\label{acknowledgments}
\addcontentsline{toc}{section}{Acknowledgments}

We thank the organizers of the NFL Big Data Bowl 2025 for releasing
publicly-available player tracking data, as well as participants at the
2025 CMSAC Football Analytics Workshop for their valuable feedback. On a
personal note, QN failed to convince RY to name the turn angle
variability measure STRAIN (2025 version), where STRAIN stands for
Shiftiness, Turning, Redirection and Agility INdex.

\section*{Data availability}\label{data-availability}
\addcontentsline{toc}{section}{Data availability}

The data provided by the NFL Big Data Bowl 2025 are available at
\url{https://www.kaggle.com/competitions/nfl-big-data-bowl-2025/data}.
Our results can be reproduced using the code provided at
\url{https://github.com/qntkhvn/turn-angle}.

\renewcommand\refname{References}
\bibliography{references.bib}

\begin{thebibliography}{}

\bibitem[Baumer et~al., 2023]{baumer2023big}
Baumer, B.~S., Matthews, G.~J., and Nguyen, Q. (2023).
\newblock Big ideas in sports analytics and statistical tools for their investigation.
\newblock {\em WIREs Computational Statistics}, 15(6):e1612.

\bibitem[Brughelli et~al., 2008]{brughelli2008understanding}
Brughelli, M., Cronin, J., Levin, G., and Chaouachi, A. (2008).
\newblock Understanding change of direction ability in sport: A review of resistance training studies.
\newblock {\em Sports Medicine}, 38(12):1045--1063.

\bibitem[B\"{u}rkner, 2017]{burkner2017brms}
B\"{u}rkner, P.-C. (2017).
\newblock {brms: An R Package for Bayesian Multilevel Models Using Stan}.
\newblock {\em Journal of Statistical Software}, 80(1):1--28.

\bibitem[Carpenter et~al., 2017]{carpenter2017stan}
Carpenter, B., Gelman, A., Hoffman, M.~D., Lee, D., Goodrich, B., Betancourt, M., Brubaker, M., Guo, J., Li, P., and Riddell, A. (2017).
\newblock Stan: A probabilistic programming language.
\newblock {\em Journal of Statistical Software}, 76(1):1--32.

\bibitem[Chu et~al., 2020]{chu2020route}
Chu, D., Reyers, M., Thomson, J., and Wu, L.~Y. (2020).
\newblock {Route identification in the National Football League}.
\newblock {\em Journal of Quantitative Analysis in Sports}, 16(2):121--132.

\bibitem[Deshpande and Evans, 2020]{deshpande2020expected}
Deshpande, S.~K. and Evans, K. (2020).
\newblock Expected hypothetical completion probability.
\newblock {\em Journal of Quantitative Analysis in Sports}, 16(2):85--94.

\bibitem[Durkin, 2019]{durkin2019clock}
Durkin, D. (2019).
\newblock {On the clock: Justice Hill’s explosive speed makes him an intriguing prospect for the Bears}.
\newblock The Athletic.
\newblock \url{https://www.nytimes.com/athletic/931228/2019/04/18/on-the-clock-justice-hills-explosive-speed-makes-him-an-intriguing-prospect-for-the-bears}.

\bibitem[Dutta et~al., 2020]{dutta2020unsupervised}
Dutta, R., Yurko, R., and Ventura, S.~L. (2020).
\newblock Unsupervised methods for identifying pass coverage among defensive backs with nfl player tracking data.
\newblock {\em Journal of Quantitative Analysis in Sports}, 16(2):143--161.

\bibitem[Gelman, 2006]{gelman2006prior}
Gelman, A. (2006).
\newblock Prior distributions for variance parameters in hierarchical models.
\newblock {\em Bayesian Analysis}, 1(3):515--534.

\bibitem[Gelman et~al., 2013]{gelman2013bayesian}
Gelman, A., Carlin, J.~B., Stern, H.~S., Dunson, D.~B., Vehtari, A., and Rubin, D.~B. (2013).
\newblock {\em Bayesian {Data} {Analysis}, {Third} {Edition}}.
\newblock Chapman \& {Hall}/{CRC} {Texts} in {Statistical} {Science}. Taylor \& Francis.

\bibitem[Gelman and Rubin, 1992]{gelman1992inference}
Gelman, A. and Rubin, D.~B. (1992).
\newblock Inference from iterative simulation using multiple sequences.
\newblock {\em Statistical Science}, 7(4):457--472.

\bibitem[Giles et~al., 2019]{giles2019machine}
Giles, B., Kovalchik, S., and Reid, M. (2019).
\newblock A machine learning approach for automatic detection and classification of changes of direction from player tracking data in professional tennis.
\newblock {\em Journal of Sports Sciences}, 38(1):106--113.

\bibitem[Giles et~al., 2024]{giles2024quantifying}
Giles, B., Peeling, P., and Reid, M. (2024).
\newblock Quantifying change of direction movement demands in professional tennis matchplay: An analysis from the australian open grand slam.
\newblock {\em Journal of Strength and Conditioning Research}, 38(3):517--525.

\bibitem[Ho and Carl, 2025]{ho2025nflreadr}
Ho, T. and Carl, S. (2025).
\newblock {\em nflreadr: Download 'nflverse' Data}.
\newblock R package version 1.4.1.07.

\bibitem[Hoffman and Gelman, 2014]{hoffman2014nuts}
Hoffman, M.~D. and Gelman, A. (2014).
\newblock The no-u-turn sampler: Adaptively setting path lengths in hamiltonian monte carlo.
\newblock {\em Journal of Machine Learning Research}, 15(47):1593--1623.

\bibitem[Hooten et~al., 2017]{hooten2017animal}
Hooten, M.~B., Johnson, D.~S., McClintock, B.~T., and Morales, J.~M. (2017).
\newblock {\em Animal Movement: Statistical Models for Telemetry Data}.
\newblock CRC Press.

\bibitem[Kassam, 2025]{kassam2025minding}
Kassam, K. (2025).
\newblock {Minding the College Football Analytics Gap}.
\newblock Presented at the 2025 MIT Sloan Sports Analytics Conference.
\newblock \url{https://youtu.be/RZNLXU2O0WM}.

\bibitem[Kovalchik, 2023]{kovalchik2023player}
Kovalchik, S.~A. (2023).
\newblock {Player Tracking Data in Sports}.
\newblock {\em Annual Review of Statistics and Its Application}, 10(1):677--697.

\bibitem[Kramer, 2019]{kramer2019jonathan}
Kramer, A. (2019).
\newblock {Jonathan Taylor Wants More, More, More}.
\newblock Bleacher Report.
\newblock \url{https://bleacherreport.com/articles/2854041-jonathan-taylor-wants-more-more-more}.

\bibitem[Levy, 2019]{levy2019combine}
Levy, L. (2019).
\newblock {Combine Conundrum: What To Make Of DK Metcalf's Agility Testing}.
\newblock Optimum Scouting.
\newblock \url{http://www.optimumscouting.com/news/dk-metcalf}.

\bibitem[Lopez et~al., 2024]{lopez2024nfl}
Lopez, M., Bliss, T., Blake, A., Mooney, P., and Howard, A. (2024).
\newblock {NFL Big Data Bowl 2025}.
\newblock \url{https://kaggle.com/competitions/nfl-big-data-bowl-2025}.

\bibitem[Lopez, 2020]{lopez2020bigger}
Lopez, M.~J. (2020).
\newblock {Bigger data, better questions, and a return to fourth down behavior: an introduction to a special issue on tracking datain the National Football League}.
\newblock {\em Journal of Quantitative Analysis in Sports}, 16(2):73--79.

\bibitem[Maclean, 2024]{maclean2024defensive}
Maclean, Q. (2024).
\newblock {Defensive Back Scouting: Using Pose Estimation to Measure Hip Fluidity}.
\newblock SumerSports.
\newblock \url{https://sumersports.com/the-zone/defensive-back-scouting-using-pose-estimation-to-measure-hip-fluidity}.

\bibitem[Nguyen et~al., 2025]{nguyen2025fractional}
Nguyen, Q., Jiang, R., Ellingwood, M., and Yurko, R. (2025).
\newblock Fractional tackles: leveraging player tracking data for within-play tackling evaluation in {A}merican football.
\newblock {\em Scientific Reports}, 15:2148.

\bibitem[Nguyen and Yurko, 2025]{nguyen2025multilevel}
Nguyen, Q. and Yurko, R. (2025).
\newblock A multilevel model with heterogeneous variances for snap timing in the national football league.
\newblock {\em arXiv preprint arXiv:2502.16313}.
\newblock \url{https://arxiv.org/pdf/2502.16313}.

\bibitem[Nguyen et~al., 2024]{nguyen2024here}
Nguyen, Q., Yurko, R., and Matthews, G.~J. (2024).
\newblock {Here Comes the STRAIN: Analyzing Defensive Pass Rush in American Football with Player Tracking Data}.
\newblock {\em The American Statistician}, 78(2):199--208.

\bibitem[{R Core Team}, 2025]{r2025language}
{R Core Team} (2025).
\newblock {\em R: A Language and Environment for Statistical Computing}.
\newblock R Foundation for Statistical Computing, Vienna, Austria.

\bibitem[Yurko et~al., 2020]{yurko2020going}
Yurko, R., Matano, F., Richardson, L.~F., Granered, N., Pospisil, T., Pelechrinis, K., and Ventura, S.~L. (2020).
\newblock {Going deep: models for continuous-time within-play valuation of game outcomes in American football with tracking data}.
\newblock {\em Journal of Quantitative Analysis in Sports}, 16(2):163--182.

\bibitem[Yurko et~al., 2024]{yurko2024nfl}
Yurko, R., Nguyen, Q., and Pelechrinis, K. (2024).
\newblock {NFL} {G}hosts: {A} framework for evaluating defender positioning with conditional density estimation.
\newblock {\em arXiv preprint arXiv:2406.17220}.
\newblock \url{https://arxiv.org/pdf/2406.17220}.

\end{thebibliography}

\end{document}